\documentclass[aps,prl,preprint,showpacs]{revtex4}
\usepackage{amssymb}
\usepackage{graphicx}
\usepackage{amsmath}
\usepackage{graphicx}

\begin{document}

\title{Universal angular magnetoresistance and spin torque\\
in ferromagnetic/normal metal hybrids}
\author{Gerrit E. W. Bauer,$^{1}$ Yaroslav Tserkovnyak,$^{2}$ Daniel
Huertas-Hernando,$^{1}$ and Arne Brataas$^{2}$ }
\affiliation{$^{1}$Department of Applied Physics and DIMES, Delft University of
Technology, 2628 CJ Delft, The Netherlands}
\affiliation{$^{2}$Harvard University, Lyman Laboratory of Physics, Cambridge,
Massachusetts 02138 USA}

\begin{abstract}
The electrical resistance of ferromagnetic/normal-metal (F/N)
heterostructures depends on the nature of the junctions which may be tunnel
barriers, point contacts, or intermetallic interfaces. For all junction
types, the resistance of disordered F/N/F
perpendicular spin valves as a function of the angle between magnetization
vectors is shown to obey a simple universal law. The spin-current induced
magnetization torque can be measured by the angular magnetoresistance of
these spin valves. The results are generalized to arbitrary
magnetoelectronic circuits.
\end{abstract}

\date{\today }
\maketitle

Magnetoelectronics achieves new functionalities by incorporating
ferromagnetic materials into electronic circuits. The giant
magnetoresistance, \textit{i.e.} the dependence of the electrical resistance
on the relative orientation of the magnetizations of two ferromagnets in a
ferromagnetic/normal/ferromagnetic (F/N/F)
metal structure or \textquotedblleft spin valve\textquotedblright ,\ is
applied in read heads of high information density magnetic storage systems 
\cite{GMR}. Usually, such a device is viewed as a single bit, the
magnetizations vectors being either parallel or antiparallel. Early seminal
contributions by Slonczewski \cite{Slon96} and Berger \cite{Berger96}
revealed fundamentally new physics and technological possibilities of
noncollinearity, which triggered a large number of experimental and
theoretical research. An important example is the non-equilibrium
spin-current induced torque (briefly, spin torque) which one ferromagnet can
exert on the magnetization vector of a second magnet through a normal metal.
This torque can be large enough to dynamically turn magnetizations \cite%
{Tsoi98}, which is potentially interesting as a low-power switching
mechanism for magnetic random access memories \cite{Inomata01}. The spin
torque is also essential for novel magnetic devices like the spin-flip
transistor \cite{Brataas00,SFT,Xia02}, detection of spin-precession \cite%
{Huertas00}, the Gilbert damping of the magnetization dynamics in thin
magnetic films \cite{Tserkovnyak:prl02}, and spin-injection induced by
ferromagnetic resonance \cite{Brataas02}.

Recently, two theoretical approaches have been developed which address
charge and spin transport in diffusive noncollinear magnetic hybrid
structures. The magnetoelectronic \textquotedblleft circuit
theory\textquotedblright\ \cite{Brataas00} is based on the division of the
system into discrete resistive elements over which the applied potential
drops, and low-resistance nodes at quasi-equilibrium (as in Fig.~1(a)). The
electrical properties are then governed by generalized Kirchhoff rules in
Pauli spin space and can be computed easily. Each resistor is thereby
characterized by four material parameters, the spin-up and spin-down
conductances $g_{\uparrow (\downarrow )}=\sum_{nm}[\delta
_{nm}-|r_{nm}^{\uparrow (\downarrow )}|^{2}]$ as known from the scattering
theory of transport \cite{Datta95}, as well as the real and imaginary part
of the \textquotedblleft mixing conductance\textquotedblright\ $g^{\uparrow
\downarrow }=\sum_{nm}[\delta _{nm}-r_{nm}^{\uparrow }(r_{nm}^{\downarrow
})^{\ast }],$ where $r_{nm}^{s}$ is the reflection coefficient between $n-$%
th and $m-$th transverse modes of an electron with spin $s$ in the normal
metal at the contact to a ferromagnet. Waintal \textit{et al. }\cite%
{Waintal00} studied the random matrix theory of transport in noncollinear
magnetic systems as sketched in Fig.~1(b). Their formalism did not require
the assumption of highly resistive elements, but the algebra of the $4\times
4$ scattering matrices in spin space seemed so complex, that analytical
results were obtained in limiting cases only.

Both theories are not valid in the limit of intermetallic interfaces in a
diffuse enviroment (see Fig.~1c) like the perpendicular spin valves, studied
thoroughly by the Michigan State University collaboration \cite{Pratt1} and
others \cite{Gijs,GMRREV}. These studies provided a large body of evidence
for the two-channel (\textit{i.e.} spin-up and spin-down) series resistor
model and a wealth of accurate transport parameters like the interface
resistances for various material combinations. Transport through transparent
interfaces in a diffuse environment has been studied for \emph{collinear}
magnetizations by Schep \textit{et al}. \cite{Schep97}. Under the condition
of isotropy of scattering by disorder, it was found that the bulk
resistances, which are proportional to the layer thicknesses, are in series
with interface resistances, for each spin $s$ 
\begin{equation}
\frac{1}{\tilde{g}_{s}}=\frac{1}{g_{s}}-\frac{1}{2}\left( \frac{1}{N_{s}^{F}}%
+\frac{1}{N_{N}}\right) ,  \label{Schep}
\end{equation}%
where $N_{s}^{F}$ and $N_{N}$ are the number of modes of the bulk materials
on both sides of the F/N contact. Physically, in Eq.~(\ref%
{Schep}) a spurious Sharvin resistance is substracted from the result of
scattering theory. This correction is large for transparent interfaces and
essential to obtain agreement between experimental results and
first-principles calculations \cite{Schep97,Stiles00,Xia01}.

In exchange-biased spin valves, it is possible to measure the electric
resistance as a function of the angle between magnetizations, which has been
analyzed experimentally and theoretically \cite%
{Dauguet:prb96,Vedyayev,Giacomo}. The present study has been motivated by
Pratt's observation that this angular magnetoresistance could accurately be
fitted by the form \cite{Brataas00} 
\begin{equation}
\frac{R\left( \theta \right) -R\left( 0\right) }{R\left( \pi \right)
-R\left( 0\right) }=\frac{1-\cos \theta }{\chi \left( 1+\cos \theta \right)
+2}  \label{AMR}
\end{equation}%
with one free parameter $\chi $ that is given by circuit theory 
\begin{equation}
\chi =\frac{1}{1-p^{2}}\frac{\left\vert \eta \right\vert ^{2}}{ {\rm Re}\eta 
}-1
\end{equation}%
in terms of the normalized mixing conductance $\eta =2g_{\uparrow \downarrow
}/g$, the polarization $p=\left( g_{\uparrow }-g_{\downarrow }\right) /g$,
and the average conductance $g=g_{\uparrow }+g_{\downarrow }$. This was
surprising, since the circuit theory, as mentioned above, was not designed
for metallic multilayers, and, indeed, the numerical value of fitted
parameters did not make sense, also after including effects of bulk
scattering in the ferromagnetic layers \cite{Huertas02}.

In the following we develop a theory of transport in disordered
magnetoelectronic circuits and devices in the diffuse regime which unifies
and extends previous theoretical approaches. We find simple analytical
results with parameters that are accessible to realistic
electronic-structure calculations. The angular magnetoresistance for
perpendicular spin valves agrees with the universal form [Eq.~(\ref{AMR})]
in agreement with measurements \cite{Giacomo}, and is used to determine the
mixing conductance and spin torque. The theory is valid under two
conditions: (i) the system should be diffusive, \textit{i.e.} the elastic
mean free path $\ell $ (including scattering at interfaces) should be
smaller than typical sample scales and (ii) the ferromagnetic elements
should have an exchange splitting $\Delta ,$ which is large enough that the
magnetic coherence length $\ell _{c}=\hbar /\sqrt{2m\Delta }<\min \left(
\ell ,d_{F}\right) ,$ where $d_{F}$\ is the thickness of the ferromagnetic
layer. These conditions are usually fulfilled in transition-metal systems:
Deviations from diffusive behavior, like quantum-size effects and breakdown
of the series resistor model, are small or controversial \cite%
{Xia01,deviations}, whereas the magnetic coherence length is of the same
order as the lattice constant in high-$T_{c}$ transition-metal ferromagnets 
\cite{Tserkovnyak:prl02,Zangwill}. We obtain identical results by two
methods: The first one is a combination of the Boltzmann-like method of
Schep \textit{et al.} \cite{Schep97} for collinear systems and the
random-matrix theory of Waintal \textit{et al.} \cite{Waintal00}. The second
one is an extension of magnetoelectronic circuit theory \cite{Brataas00} to
arbitrary resistors.

Let us consider planar spin-valve structures as shown in Fig.~1. We assume
the existence of a distribution function at a certain position $x$ in the
sample (a \textquotedblleft node\textquotedblright ), which in
spin-polarized systems has 8 elements $f_{ss^{\prime }}^{\pm }\left(
x\right) $. We arrange them into a $4\times 1$ vector $\vec{f}^{\pm }=\left(
f_{\uparrow \uparrow }^{\pm },f_{\uparrow \downarrow }^{\pm },f_{\downarrow
\uparrow }^{\pm },f_{\downarrow \downarrow }^{\pm }\right) ^{T}$ as well as
into a $2\times 2$ matrix, denoted by a hat:%
\begin{equation}
\hat{f}^{\pm }\left( x\right) =\left( 
\begin{array}{cc}
f_{\uparrow \uparrow }^{\pm }\left( x\right) & f_{\downarrow \uparrow }^{\pm
}\left( x\right) \\ 
f_{\uparrow \downarrow }^{\pm }\left( x\right) & f_{\downarrow \downarrow
}^{\pm }\left( x\right)%
\end{array}%
\right) .
\end{equation}%
The superscript denotes that the distribution is in general anisotropic in
reciprocal space, $+$ for right-moving $-$ for left moving, indicating that,
in contrast to \cite{Brataas00,Waintal00}, the current density in the nodes
is not negligible. The distribution functions at different nodes are matched 
\textit{via} boundary conditions: 
\begin{subequations}
\label{bound}
\begin{eqnarray}
\vec{f}^{+}\left( x_{B}\right) &=&\check{T}_{A\rightarrow B}\vec{f}%
^{+}\left( x_{A}\right) +\check{R}_{B\rightarrow B}\vec{f}^{-}\left(
x_{B}\right) \\
\vec{f}^{-}\left( x_{A}\right) &=&\check{R}_{A\rightarrow A}\vec{f}%
^{+}\left( x_{A}\right) +\check{T}_{B\rightarrow A}\vec{f}^{-}\left(
x_{B}\right) ,
\end{eqnarray}%
where the $4\times 4$ transmission and reflection probability matrices
(indicated by the caret) have elements like \cite{Waintal00}: 
\end{subequations}
\begin{equation}
\left[ \check{T}_{A\rightarrow B}\right] _{ij}=\frac{1}{N_{i}^{B}}%
\sum_{nm}\left( \vec{t}_{nm}^{A\rightarrow B}\right) _{i}\left( \vec{t}%
_{nm}^{A\rightarrow B}\right) _{j}^{\dagger }
\end{equation}%
where $N_{i}^{B}=N_{\uparrow }^{B}\left( \delta _{i,1}+\delta _{i,2}\right)
+N_{\downarrow }^{B}\left( \delta _{i,3}+\delta _{i,4}\right) ,$ $N_{s}^{B}$
is the number of modes for spin $s$ in $B$, and $\vec{t}_{nm}^{A\rightarrow
B}$ is a vector of the transmission coefficients in spin space.

Let us calculate the electrical charge current in a symmetric two-terminal
spin valve with relative magnetization angle $\theta $ (Fig. 1). $x_{L}$ and 
$x_{R}$ are within left and right ferromagnets at a distance from the
interface equal to the spin diffusion length in the ferromagnet $\ell
_{sd}^{F}\gg $ $\ell _{c}$, and thus define the magnetically active region.
In the coordinate systems defined by the magnetization directions, the
transverse components of the spin accumulation in the ferromagnets vanish 
\cite{Brataas00,Zangwill} and the distributions in the magnets depend on the
local spin current densities $\gamma _{s}$ and (spin-independent) chemical
potentials $\mu $ only: 
\begin{equation}
\vec{f}^{\pm }\left( x\right) =\left( \left( \pm \gamma _{\uparrow }+\mu
\right) \left( x\right) ,0,0,\left( \pm \gamma _{\downarrow }+\mu \right)
\left( x\right) \right) .  \label{fF}
\end{equation}%
In symmetric junctions the spin current is symmetric as well, $\gamma
_{s}\left( x_{L}\right) =\gamma _{s}\left( x_{R}\right) $. The charge
current $i_{c}=\left( e^{2}/h\right) \sum_{s}N_{s}^{F}\gamma _{s}\ $divided
by the chemical potential drop equals the electrical conductance $%
G=i_{c}/\Delta \mu $. Eqs. (\ref{bound},\ref{fF}) then lead to: 
\begin{equation}
G=\frac{2e^{2}}{h}\sum_{\substack{ i=1,4  \\ j=1,4}}\left\{ N_{i}^{F}\left[ 
\check{1}-\check{T}_{L\rightarrow R}+\check{R}_{R\rightarrow R}\right] ^{-1}%
\check{T}_{L\rightarrow R}\right\} _{ij}\,.  \label{GSV}
\end{equation}%
In principle, the matrices $\check{T}$ and $\check{R}$ do not need to be
approximate.

In dirty systems, more nodes may be introduced at convenient locations in
the sample and Eqs.~(\ref{bound}) implies that total transport probability
matrices can be composed in terms of those of individual elements by
semiclassical concatenation rules \cite{Shapiro}. For instance, the
transmission through a F$\left( 0\right) $/N/F$\left( \theta \right) $
double heterojunction as in Fig.~1 (without bulk scattering) takes the form: 
\begin{equation}
\check{T}\left( \theta \right) \equiv \check{T}_{\mathrm{N\rightarrow F}%
}\left( \theta \right) \left[ \check{1}-\check{R}_{\mathrm{N\rightarrow N}%
}\left( 0\right) \check{R}_{\mathrm{N\rightarrow N}}\left( \theta \right) %
\right] ^{-1}\check{T}_{\mathrm{F\rightarrow N}}\left( 0\right) .
\end{equation}%
These rules have been derived from the (phase-coherent) scattering theory by
averaging over random matrices \cite{Waintal00} and found to be valid to
leading order in $N_{N}^{-1}$, where $N_{N}$ is the number of transport
channels in the normal metal. Bulk impurity scattering can be represented by
diagonal matrices \cite{Schep97,Waintal00}%
\begin{equation}
\left( \check{T}_{B}\right) _{ss^{\prime }}=\left( 1+\frac{1}{N_{s}^{B}}+%
\frac{e^{2}}{h}\frac{\rho _{s}^{B}d_{B}}{A_{B}}\right) ^{-1}\delta
_{ss^{\prime }}\ 
\end{equation}%
where $\rho _{s}^{B},$ $d_{B},$ $A_{B}$ are the bulk resistivities,
thickness and cross section of the bulk material $B$, repectively.

The problem can be simplified by transformations into the coordinate systems
defined by the magnetization directions of the ferromagnets. In terms of the
spin-rotation 
\begin{equation}
\hat{U}=\left( 
\begin{array}{cc}
\cos \frac{\theta }{2} & -\sin \frac{\theta }{2} \\ 
\sin \frac{\theta }{2} & \cos \frac{\theta }{2}%
\end{array}%
\right)
\end{equation}%
and projection matrices $\left( s=\pm 1\right) $%
\begin{equation}
\hat{u}_{s}\left( \theta \right) =\frac{1}{2}\left( 
\begin{array}{cc}
1+s\cos \theta & s\sin \theta \\ 
s\sin \theta & 1-s\cos \theta%
\end{array}%
\right) ,  \label{us}
\end{equation}%
the interface scattering matrices (omitting the mode indices for simplicity)
are transformed as \cite{Brataas00} $t_{ss^{\prime }}^{F\rightarrow
N}=U_{ss^{\prime }}t_{s^{\prime }}^{cF},$ $t_{ss^{\prime }}^{N\rightarrow
F}=t_{s}^{cN}U_{ss^{\prime }}^{\dagger }$ $\hat{r}_{N\rightarrow N}=\sum_{s}%
\hat{u}_{s}r_{s}^{cN}$, and $r_{ss^{\prime }}^{F\rightarrow
F}=r_{s}^{cF}\delta _{ss^{\prime }}$, where the superscript $c$ indicates
that the matrices should be evaluated in the reference frame of the local
magnetization and spin-flip scattering in the contacts has been disregarded.

The angular magnetoresistance can now be evaluated analytically for our spin
valve in terms of the three interface conductances $g_{\uparrow },$ $%
g_{\downarrow }$, $g_{\downarrow \uparrow }$ defined above, the bulk number
of modes $N_{s}^{F},$ $N^{N}$, and bulk resistances $\rho _{s}^{F},$ $\rho
^{N},$ whereas the magnetization angle and layer thicknesses are the
variables. Surprisingly, the form Eq.~(\ref{AMR}) is recovered, but with
renormalized parameters. The spin-dependent interfaces conductances are
identical to Eq.~(\ref{Schep}), whereas, including the effect of bulk
scattering,%
\begin{equation}
\frac{1}{\tilde{g}_{\uparrow \downarrow }}=\frac{1}{g_{\uparrow \downarrow }}%
+\frac{1}{2}\left( \frac{\rho _{N}d_{N}}{A}-\frac{1}{N_{N}}\right) .
\label{rmix}
\end{equation}%
By letting $N_{s}^{F}\rightarrow \infty $ we are in the regime of \cite%
{Waintal00}. The circuit theory is recovered when, additionally, $%
N_{N}\rightarrow \infty .$ The bare mixing conductance is bounded not only
from below $ {\rm Re}g_{\uparrow \downarrow }\geqslant g/2$\ \cite{Brataas00}%
, but also from above $\left\vert g_{\uparrow \downarrow }\right\vert ^{2}/%
{\rm Re}g_{\uparrow \downarrow }\leqslant 2N_{N}.$ The polarization and
relative mixing conductances are also renormalized, with $0<\left\vert 
\tilde{\eta}\right\vert <\infty .$

It is not obvious how these results should be generalized to more
complicated circuits and devices and to the presence of spin-flip scattering
in the normal metal. The magnetoelectronic circuit theory \cite{Brataas00}
does not suffer from these drawbacks. In the following we demonstrate that
above results can be obtained with less effort, proving that with the
renormalization of the transport parameters by subtracting Sharvin
resistances, circuit theory remains valid for arbitrary contacts. To this
end we construct the fictitious circuit depicted in Fig.~2. Consider a
junction which in conventional circuit theory is characterized by a matrix
conductance $\hat{g}$ leading to a matrix current $\hat{\imath}$ when the
normal and ferromagnetic distributions $\hat{f}_{L}$ and $\hat{f}_{R}$ are
not equal. When the distributions of the nodes are isotropic, we know from
circuit theory that:%
\begin{equation}
\hat{\imath}=\sum_{ss^{\prime }}\left( \hat{g}\right) _{ss^{\prime }}\hat{u}%
_{s}\left( \hat{f}_{L}-\hat{f}_{R}\right) \hat{u}_{s^{\prime }}\,,
\end{equation}%
where the projection matrices $\hat{u}_{s}$ are defined in Eq.~(\ref{us})
and {$(\hat{g})_{ss}=g_{s},$ $(\hat{g})_{s,-s}=$}$g_{s,-s}$. Introducing
lead conductances, which modify the distributions $\hat{f}_{L}\rightarrow 
\hat{f}_{1}$ and $\hat{f}_{2}\leftarrow \hat{f}_{R},$ respectively, we may
define a (renormalized) conductance matrix 
$\hat{\tilde{g}}$%
, which causes an identical current $\hat{\imath}$ for the reduced (matrix)
potential drop:%
\begin{equation}
\hat{\imath}=\sum_{ss^{\prime }}\left( 
\hat{\tilde{g}}%
\right) _{ss^{\prime }}\hat{u}_{s}\left( \hat{f}_{1}-\hat{f}_{2}\right) \hat{%
u}_{s^{\prime }}\,.  \label{i1}
\end{equation}%
When the lead conductances are now chosen to be one-half of the Sharvin
conductances, and using (matrix) current conservation: 
\begin{eqnarray}
\hat{\imath} &=&2N_{N}\left( \hat{f}_{L}-\hat{f}_{1}\right)  \label{i2} \\
&=&\sum_{s}2N_{s}^{F}\hat{u}_{s}\left( \hat{f}_{2}-\hat{f}_{R}\right) \hat{u}%
_{s},
\end{eqnarray}%
straightforward matrix algebra leads to the result that 
$\hat{\tilde{g}}$
is identical to the renormalized interface conductances found above [Eqs.~(%
\ref{Schep}) and, without the bulk term, (\ref{rmix})]. By replacing $\hat{g}
$ by 
$\hat{\tilde{g}}$
we not only recover results for the spin valve obtained above, but we can
now use the renormalized parameters also for circuits with arbitrary
complexity and transparency of the contacts. Also spin-flip scattering in N
can be included \cite{Brataas00}; it does not affect the form of Eq.~(\ref%
{AMR}) either, but only reduces the parameter $\tilde{\chi}.$

Experimental values for the parameters for Cu/Permalloy (Py) spin valves are 
$\tilde{\chi}=1.2$ and $\tilde{p}=0.6$ \cite{Giacomo}. Disregarding a very
small imaginary component of the mixing conductance \cite{Xia02}, using the
known values for the bulk resistivities, the theoretical Sharvin resistance
for Cu ($0.55\cdot 10^{15\;}\mathrm{\Omega }^{-1}\mathrm{m}^{-2}$/spin \cite%
{Schep97}), and the spin-flip length of Py as the effective thickness of the
ferromagnet $\left( \ell _{sd}^{F}=5\text{ nm}\ \text{\cite{Pratt1}}\right) $%
, we arrive at the bare Cu/Py interface mixing conductance $G_{\uparrow
\downarrow }=0.39\left( 3\right) \cdot 10^{15\;}\mathrm{\Omega }^{-1}\mathrm{%
m}^{-2}$, which is close to that of Co/Cu \cite{Xia02}.

The analytical expression for the spin torque on either ferromagnet, \textit{%
i.e.} the spin current normal to the magnetization direction, reads%
\begin{equation}
L\left( \theta \right) =\frac{\tilde{p}\tilde{g}}{2}\frac{\tilde{\eta}\sin
\theta }{\left( \tilde{\eta}-1\right) \cos \theta +1+\tilde{\eta}}\frac{%
\Delta \mu }{2\pi },
\end{equation}%
in terms of parameters which can be measured as well as computed from first
principles. Previous results \cite{Slon96,Waintal00} are recovered in the
limit that $\tilde{\eta}\rightarrow 2$ and $\tilde{p}\rightarrow 1.$ By the
generalized circuit theory it is straightforward to compute the torque on
the base contact of the spin-flip transistor with antiparallel source-drain
magnetizations (three identical contacts) \cite{Xia02}. Interestingly, it is
larger and has a symmetric and flatter dependence on the angle of the base
magnetization direction $\theta :$ 
\begin{equation*}
L_{b}\left( \theta \right) =\frac{\tilde{p}\tilde{g}\tilde{\eta}\sin \theta 
}{\left( 1-\tilde{\eta}\right) \cos ^{2}\theta +2+\tilde{\eta}}\frac{\Delta
\mu }{2\pi }.
\end{equation*}%
This opens the way to engineer materials and device configurations to
optimize switching properties of magnetic random access memories.

We would like to thank Bill Pratt for attracting our attention to the
problem and sharing unpublished experimental data. We acknowledge
discussions with Paul Kelly, Ke Xia, Jack Bass, Bert. I Halperin, Yuli
Nazarov, as well as support by FOM, the Schlumberger Foundation, DARPA Award
No. MDA 972-01-1-0024, NSF Grant NO. DMR 99-81283 and the NEDO joint
research program \textquotedblleft Nano-Scale
Magnetoelectronics\textquotedblright . G.B. is grateful for the hospitality
of Dr. Y. Hirayama and his group at the NTT Basic Research Laboratories.

\begin{figure}[h]
\caption{Different realizations of perpendicular spin
valves. (a) Highly resistive junctions like point contacts and tunneling
barriers limit the conductance. (b) Spin valve in a geometrical constriction
amenable to the scattering theory of transport. (c) Magnetic multilayers
with transparent interfaces. $\protect\theta $ is the angle between
magnetization directions.}
\end{figure}

\begin{figure}[h]
\caption{Fictitious device which illustrates the
generalization of circuit theory to transparent resistors as discussed in
the text.}
\end{figure}


\begin{thebibliography}{99}
\bibitem{GMR} http://www.almaden.ibm.com/sst/

\bibitem{Slon96} J.C. Slonczewski, J. Magn. Magn. Mater. \textbf{159}, L1
(1996).

\bibitem{Berger96} L. Berger, Phys. Rev. B \textbf{54}, 9353 (1996).

\bibitem{Tsoi98} M. Tsoi \textit{et al.}, Phys. Rev. Lett. \textbf{80}, 4281
(1998); J.-E. Wegrowe \textit{et al.}, Europhys. Lett. \textbf{45}, 626
(1999); J.Z. Sun, J. Magn. Magn. Mater. \textbf{202}, 157 (1999); E.B. Myers 
\textit{et al.}, Science \textbf{285}, 867(1999); J.A. Katine \textit{et al.}%
, Phys. Rev. Lett. \textbf{84}, 3149 (2000); J. Grollier \textit{et al.},
Appl. Phys. Lett. \textbf{78}, 3663 (2001).

\bibitem{Inomata01} K. Inomata, IEICE Transactions on Electronics \textbf{%
E84-C}, 740 (2001); J.C. Slonczewski, cond-mat/0205055.

\bibitem{Brataas00} A. Brataas, Yu.V. Nazarov, and G.E.W. Bauer, Phys. Rev.
Lett. \textbf{84}, 2481 (2000); Eur. Phys. J. B \textbf{22}, 99 (2001).

\bibitem{SFT} G.E.W. Bauer, Yu.V. Nazarov, and A. Brataas, Physica E \textbf{%
74}, 137 (2001).

\bibitem{Xia02} K. Xia, P. J. Kelly, G. E. W. Bauer, A. Brataas, and I.
Turek, Phys. Rev. B \textbf{65}, 2204XX(R) (2002).

\bibitem{Huertas00} D.H. Hernando, Yu.V. Nazarov, A. Brataas, and
G.E.W. Bauer, Phys. Rev. B \textbf{62}, 5700 (2000).

\bibitem{Tserkovnyak:prl02} Y. Tserkovnyak, A. Brataas, G.E.W. Bauer, Phys.
Rev. Lett. \textbf{88}, 117601 (2002).

\bibitem{Brataas02} A. Brataas, Y. Tserkovnyak, G.E.W. Bauer, and B.
Halperin, cond-mat/0205028.

\bibitem{Datta95} S. Datta, \emph{Electronic Transport in Mesoscopic Systems}
(Cambridge University Press, Cambridge, 1995).

\bibitem{Waintal00} X. Waintal, E.B. Myers, P.W. Brouwer, and D.C. Ralph,
Phys. Rev. B \textbf{62}, 12317 (2000).

\bibitem{Pratt1} W.P. Pratt, Jr., S.-F. Lee, J.M. Slaughter, R. Loloee, P.A.
Schroeder, and J. Bass, Phys. Rev. Lett. \textbf{66}, 3060 (1991); J. Bass
and W.P. Pratt, J. Magn. Magn. Mater. \textbf{200}, 274 (1999).

\bibitem{Gijs} M.A.M. Gijs, S.K.J. Lenczowski, and J.B. Giesbers, Phys. Rev.
Lett. \textbf{70}, 3343 (1993).

\bibitem{GMRREV} For recent reviews see: M.A.M. Gijs and G.E.W. Bauer,
Advances in Physics \textbf{46}, 285 (1997); J.-P. Ansermet, J. Phys.-Cond
Mat. \textbf{10}, 6027 (1998); A. Barth\'{e}l\'{e}my, A. Fert and F.
Petroff, in \emph{Handbook of Magnetic Materials}, Vol. 12, edited by K.H.J.
Buschow (1999); E. Tsymbal and D. G. Pettifor, Sol. State Phys. \textbf{56},
113 (2001).

\bibitem{Schep97} K.M. Schep \textit{et al}. Phys.~Rev.~B \textbf{56},
10805(1997); G.E.W. Bauer, K.M. Schep, P.J. Kelly, K. Xia, J. Phys. D, in
press.

\bibitem{Stiles00} M. D. Stiles and D. R. Penn, Phys. Rev. B \textbf{61},
3200 (2000).

\bibitem{Xia01} K. Xia, P.J. Kelly, G.E.W. Bauer, I. Turek, J. Kudrnovsk\'{y}%
, and V. Drchal, Phys. Rev. B \textbf{63}, 064407 (2001).

\bibitem{Dauguet:prb96} P. Dauguet \textit{et al}., Phys. Rev. B \textbf{54}%
, 1083 (1996).

\bibitem{Vedyayev} A. Vedyayev, N. Ryzhanova, B. Dieny, P. Dauguet, P.
Gandit, and J. Chaussy Phys. Rev. B 55, 3728 (1997).

\bibitem{Giacomo} L. Giacomoni, B. Dieny, W.P. Pratt, Jr., R. Loloee and M.
Tsoi, to be published.

\bibitem{Huertas02} D.H. Hernando, G.E.W. Bauer, Yu.V. Nazarov, J.
Mag. Mag. Mat., in press.

\bibitem{deviations} E.Y. Tsymbal, Phys. Rev. B \textbf{62}, R3608 (2000);
D. Bozec \textit{et al.}, Phys. Rev. Lett. 85, 1314 (2000); A. Shpiro and
P.M. Levy, Phys. Rev. B \textbf{63}, 014419 (2001); K. Eid \textit{et al.}
Phys. Rev. B 65, 054424 (2002).

\bibitem{Zangwill} M.D. Stiles and A. Zangwill, cond-mat/0202397.

\bibitem{Shapiro} B. Shapiro, Phys. Rev. B \textbf{35,} 8256\ (1987); M.
Cahay, M. McLennan, and S. Datta, Phys. Rev. B \textbf{37}, 10125 (1988); A.
Brataas and G.E.W. Bauer, Phys. Rev. B \textbf{49}, 14684(1994).
\end{thebibliography}
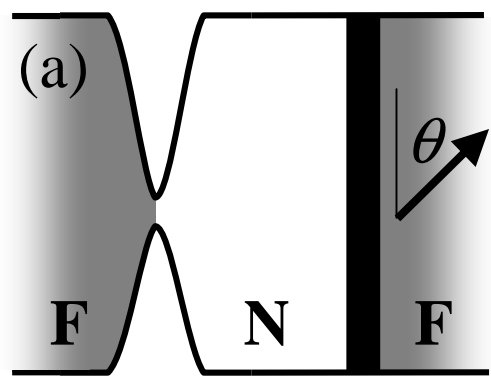
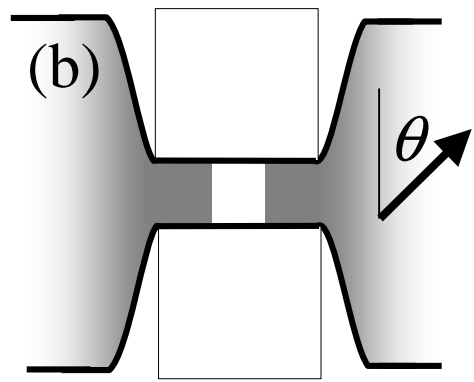
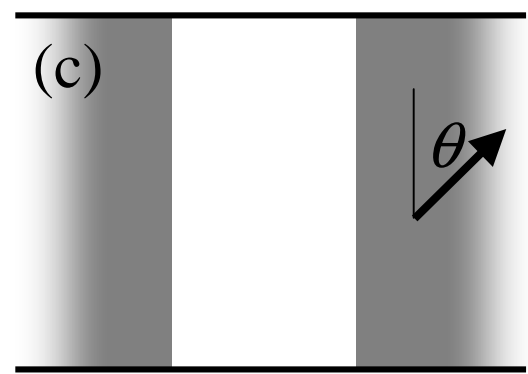
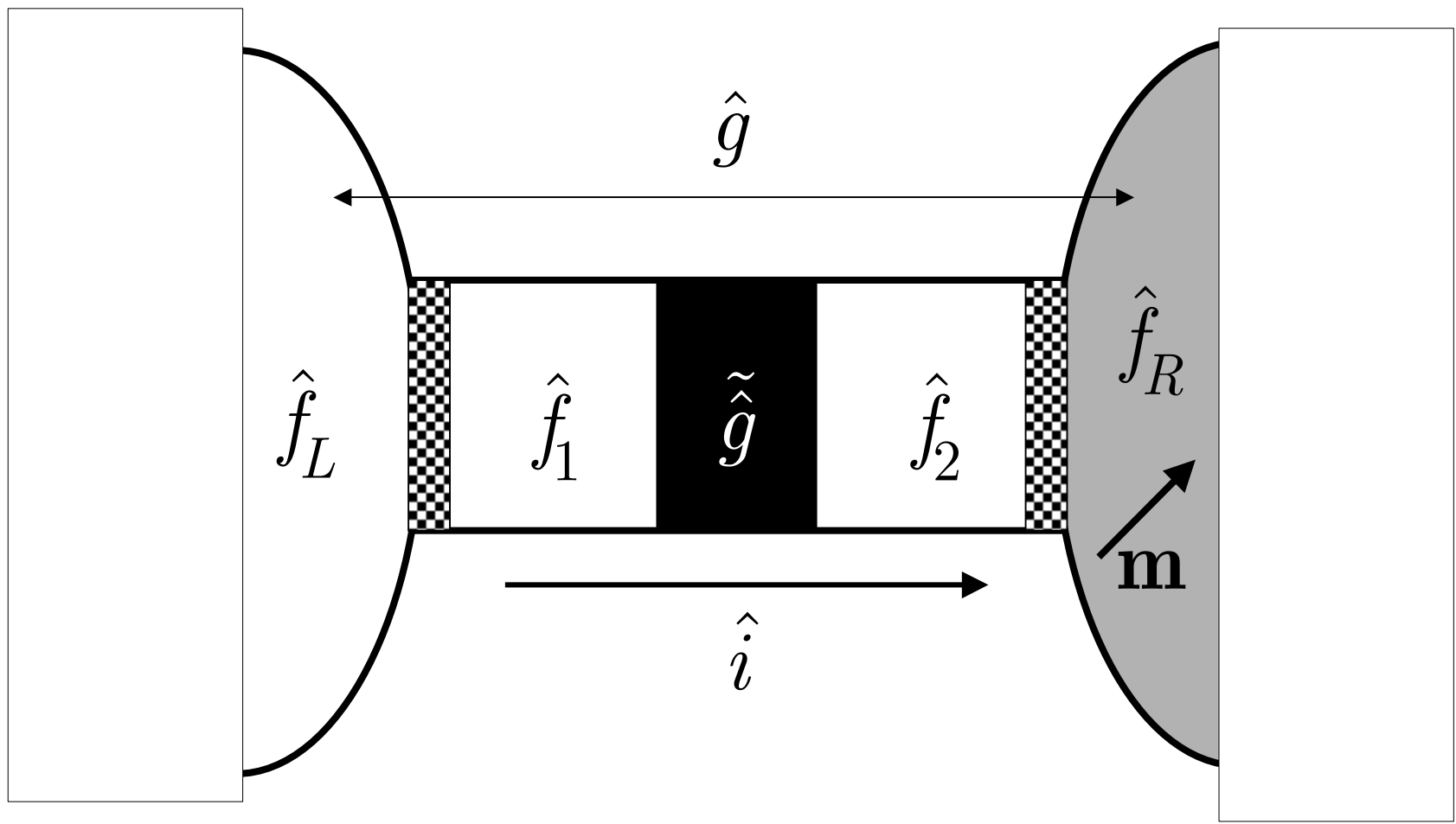
\end{document}